\documentclass[preprint,12pt,authoryear]{elsarticle-for-tr}

\usepackage{amssymb}
\usepackage[utf8]{inputenc}
\usepackage[T1]{fontenc}
\usepackage[english,lined,ruled,boxed,commentsnumbered,linesnumbered,noend]{algorithm2e}
\usepackage{mathtools}
\usepackage{comment}
\usepackage{subfig}
\usepackage{graphicx}
\usepackage{caption}
\usepackage{float}
\usepackage{tablefootnote}
\usepackage{threeparttable}
\usepackage[nameinlink]{cleveref}
\Crefname{figure}{}{}

\begin{document}

\begin{frontmatter}




\title{A flexible algorithm to offload DAG applications for edge computing}


\cortext[cor]{Corresponding author.}
\author[inst1,inst2]{Gabriel F. C. de Queiroz\corref{cor}}
\ead{gabriel.queiroz@cefet-rj.br}

\affiliation[inst1]{organization={Coordenação do Curso de Engenharia de Telecomunicações,\\Centro Federal de Educação Tecnológica Celso Suckow da Fonseca},
            addressline={\\Rua General Canabarro, 485, Sala E-204}, 
            city={Rio de Janeiro},
            postcode={20271-110}, 
            state={RJ},
            country={Brazil}}

\author[inst2]{José F. de Rezende}
\author[inst2]{Valmir C. Barbosa}

\affiliation[inst2]{organization={Programa de Engenharia de Sistemas e Computa\c c\~ao, COPPE,\\Universidade Federal do Rio de Janeiro},
            addressline={\\Centro de Tecnologia, Sala H-319}, 
            city={Rio de Janeiro},
            postcode={21941-972}, 
            state={RJ},
            country={Brazil}}

\begin{abstract}
Multi-access Edge Computing~(MEC) is an enabling technology to
leverage new network applications, such as virtual/augmented reality, by
providing faster task processing at the network edge. This is
done by deploying servers closer to the end users to
run the network applications. These applications are often intensive in
terms of task processing, memory usage, and communication; thus mobile
devices may take a long time or even not be
able to run them efficiently. By transferring (offloading) the execution
of these applications to the servers at the network edge,
it is possible to achieve a lower completion time (makespan)
and meet application requirements. However, offloading multiple entire applications to
the edge server can overwhelm its hardware and communication channel,
as well as underutilize the mobile devices' hardware. In this
paper, network applications are modeled as Directed Acyclic Graphs~(DAGs) and partitioned into
tasks, and only part of these tasks are offloaded to
the edge server. This is the DAG application partitioning and
offloading problem, which is known to be NP-hard.
To approximate its solution, this paper proposes the FlexDO algorithm.
FlexDO combines a greedy phase with a permutation phase to
find a set of offloading decisions, and then chooses the
one that achieves the shortest makespan. FlexDO is compared with
a proposal from the literature and two baseline decisions, considering
realistic DAG applications extracted from the Alibaba Cluster Trace Program. Results
show that FlexDO is consistently only 3.9\% to 8.9\% above the
optimal makespan in all test scenarios, which include different levels
of CPU availability, a multi-user case, and different communication
channel transmission rates. FlexDO outperforms both baseline solutions by a wide margin, and is three times closer to the optimal
makespan than its competitor.
\end{abstract}




\end{frontmatter}


\section{Introduction}
\label{sec:1}


The rise of new network applications and services in recent years, such as mobile social networks~\citep{hu2015, huang2017}, live video streaming~\citep{aliyu2018,yaqoob2020,jedari2021}, and the metaverse~\citep{wang2022}, for example, imposes increasingly demanding communication and computing requirements on user devices in order to provide a fully connected, immersive, and low-latency experience~\citep{parvez2018}. These stringent processing, memory, and bandwidth requirements are in general not met by the mobile devices that run those applications.

Even where user devices, such as smartphones and tablets, are capable of running such applications, this can come at great cost: having a time-consuming execution, slowing down the response time to user commands, heating up the device, and depleting the battery~\citep{wu2017}. One possible solution to this problem is to transfer (offload) the applications to a high-end server at the network core~\citep{salaht2020}, away from the end user. However, latency-sensitive applications can suffer from long transmission times if the server is located too far, resulting in a low quality of experience~\citep{mao2017}.

An additional problem with relying on central cloud servers is the ever-increasing computing and communication demands of applications.~The more stringent their requirements and the number of mobile users running them, the greater the burden on the current network infrastructure and existing servers. \citet{cisco2020} estimated that nearly 300 million mobile applications will be downloaded by the end of 2023, including social media, business, and gaming. Also according to~\citet{cisco2020}, there will be over 13 billion mobile devices and connections by year end. The high number of mobile users demanding applications and data transmissions from/to cloud servers imposes a real challenge in terms of bandwidth availability, transmission times, and end-to-end delay, which are critical for latency-sensitive applications~\citep{salaht2020}.

Multi-access Edge Computing~(MEC) is thus key to enable the upcoming demand for new applications and services~\citep{shi2016,mao2017}. MEC is a networking paradigm that brings part of the computational (memory, CPU) and transmission resources available in high-end servers at the network core, such as cloud data centers, to edge servers located at the network edge, usually connected to access points and close to the user~\citep{shi2016}. However, if there are many users fully depending on a single edge server, this may slow down the edge server and yield poor results in terms of reducing the completion time, also called makespan, of a latency-sensitive application.

To lighten the load on the edge server, it is possible to use it more efficiently, letting the mobile devices run parts of the applications. Since network applications are in general not monolithic and can be characterized as a set of multiple processing parts, called tasks, they are often described as Directed Acyclic Graphs~(DAGs)~\citep{jia2014} and task computation can be divided between mobile device and edge server. When a task runs on an edge server, it is said to have been offloaded to that server. This is a known research problem in MEC~\citep{lin2020}, called the DAG Application Partitioning and Offloading~(DAPO) problem.

Many articles have already addressed this problem~\citep{jia2014,wang2016,guo2019}. However, there still remain three under-explored problems in the literature. First, CPU limitations on both mobile and edge devices, bandwidth sharing in data transmission, and task dependencies are often neglected~\citep{salaht2020}. Second, realistic data for DAG applications with more complex structures~\citep{zhang2013} including more than a dozen tasks~\citep{yang2020,an2022} are rarely found in the literature, and even the proposed solutions found in the literature fit only specific scenarios~\citep{salaht2020}. Finally, the so-called one-climb policy~\citep{zhang2012,yang2016,an2022} is always taken as being true. This policy says that the optimal offloading decision should never have non-offloaded tasks between offloaded tasks. In this paper, and to the best of our knowledge for the first time, this policy is shown to be not necessarily true.

This paper proposes the Flexible DAG Offloading (FlexDO) algorithm. FlexDO generates a reduced set of offloading decisions, each specifying which device processes each task of a DAG application. The offloading decisions are generated by combining two steps. In the first, a greedy approach is followed to offload some tasks. To do this, FlexDO avoids transmitting large amounts of data by offloading tasks in pairs, since there is no transmission when two tasks run on the same device. In the second step, all possible offloading decisions of the remaining tasks are generated and the one resulting in the shortest makespan for the DAG application is selected.

The main contributions of this paper are as follows.

\begin{itemize}
    \item The DAPO problem is formulated taking into account CPU and channel bandwidth sharing to calculate makespan.
    \item The one-climb policy is shown not to be a requirement for the optimal offloading decision on general DAGs, contrary to the common belief expressed in the literature. In fact, our results show that about 17\% of optimal decisions for over a thousand DAG applications do not support this belief.
    \item The FlexDO heuristic is introduced to efficiently solve the problem by providing an offloading decision in reasonable time and reducing application makespan. Our results show that FlexDO achieves makespan values only 3.9\% to 8.9\% above the optimal in all test scenarios, which include single-user and multi-user cases, as well as multiple transmission data rates.
    \item Over a thousand realistic DAG applications are extracted from the Alibaba cluster trace and used to compare the performance of FlexDO with two baseline solutions and the main competitor from the literature. FlexDO outperforms all of them.
\end{itemize}

The remainder of this paper is organized as follows. Section~\ref{sec:2} presents the state of the art of task offloading in MEC. In Section~\ref{sec:3}, the DAPO problem is defined. Section~\ref{sec:4} discusses the proposed FlexDO heuristic solution. Section~\ref{sec:5} shows how realistic DAG applications are obtained for experimentation and their composition. Our results are discussed in Section~\ref{sec:6}. Finally, Section~\ref{sec:7} concludes this paper.

\section{Related work}
\label{sec:2}


Efficiently distributing the execution of computational tasks across different devices is a common concern in many computing systems~\citep{lee2012,zhang2013,yu2016,zhang2019}. Regarding edge computing, challenges arise within the DAPO problem, such as complex application structure, application partitioning~\citep{lin2020}, shortage of computing resources in mobile and edge devices, and highly dynamic transmission conditions~\citep{shi2016}. Many surveys have already reviewed general aspects of application offloading in the Internet of Things~(IoT), as well as in multi-access edge and cloud computing~\citep{dinh2013,fernando2013,khan2014,yi2015,mach2017,mao2017,abbas2018,parvez2018,roman2018,lin2020,shakarami2020,feng2022}.

\citet{kuang2019} use Lagrangian decomposition and convex optimization to jointly reduce energy consumption and execution delay in the offloading problem for MEC applications. \citet{liao2021} propose a genetic algorithm for application offloading in a MEC-enabled ultra-dense cellular network, allowing applications to be offloaded and queued in a remote base station for sequential-only processing in order to reduce completion time and energy consumption. \citet{wang2016} investigate optimization techniques in a computation offloading problem to minimize both completion time and energy consumption, exploring dynamic voltage and frequency scaling to lower the energy consumption of mobile devices. However, these works do not consider the partitioning of applications into smaller components~\citep{kuang2019,liao2021}. Applications are offloaded in their entirety and this coarser granularity can burden the mobile device or edge server with full application executions. Even when partial offloading is considered, this is done by dividing the application into a number of bytes to be processed on the mobile device and the remainder to be processed on the edge server (as in~\citeauthor{wang2016},~\citeyear{wang2016}), but application partitioning cannot be done at any point~\citep{mao2017}, due to the structure of its internal components (tasks), state variables, and memory usage.

The following papers consider application partitioning into tasks and the data transmission between them in the offloading problem. \citet{an2022} address the offloading of sequential DAG applications and resource allocation in a MEC-assisted IoT system, reducing both the energy consumption of IoT devices and makespan under different channel conditions. \citet{duan2021} propose an optimal partitioning decision for tree-based Deep Neural Networks~(DNN) modeled as DAGs, to reduce DNN inference time of IoT applications. \citet{jia2014} propose an offloading algorithm for sequential-only or parallel-only DAG applications, exploring data transmission load balancing in a mobile cloud scenario to reduce application makespan. These publications are among those that rely on overly simplified DAG structures: sequential and parallel tasks in~\citet{zhang2013,jia2014,duan2021}, DAG applications with only a few tasks (e.g., ten tasks in~\citeauthor{an2022},~\citeyear{an2022}), and purely synthetic DAGs related to no real application in~\citet{yang2020}.

\citet{guo2019} extend the prior work in~\citet{guo2018} and is the main related work regarding the scope of the present paper, as the authors consider real-world DAG applications, CPU shortage, and communication channel capacity in the DAPO problem. They formulate a Mixed Integer Linear Programming~(MILP) problem to offload DAG applications in a MEC scenario. Then a congestion-aware heuristic is provided to solve this problem in reasonable time. Transmissions are scheduled on a First In, First Out~(FIFO) basis and upcoming transmissions may be delayed in order to avoid parallelism in multi-user scenarios. A CPU assignment scheme is also provided, with task processing being allowed only if there are CPUs available.

Despite the contributions in~\citet{guo2019}, CPU shortage is accounted for only on the edge server, thus assuming that mobile devices always have more CPUs than executable tasks at any moment. Another issue is that their algorithm gives up trying to offload when, at some point, the calculated makespan is greater than the makespan of the solution that runs entirely on the mobile device (without offloading). This can hinder the efficiency of their algorithm, because makespan reduction is not always monotonic.

This paper aims to solve the DAPO problem while considering the dependencies between tasks, the limitations of channel capacity for parallel transmissions, and the limitations of concurrent task processing with scarce CPUs. This is done in order to reduce the completion time (makespan) of real DAG applications taken from a data set with more complex structures than those found in the literature.

\section{Problem definition}
\label{sec:3}

\begin{figure}[t]
  \centering
  \includegraphics[width=0.7\textwidth]{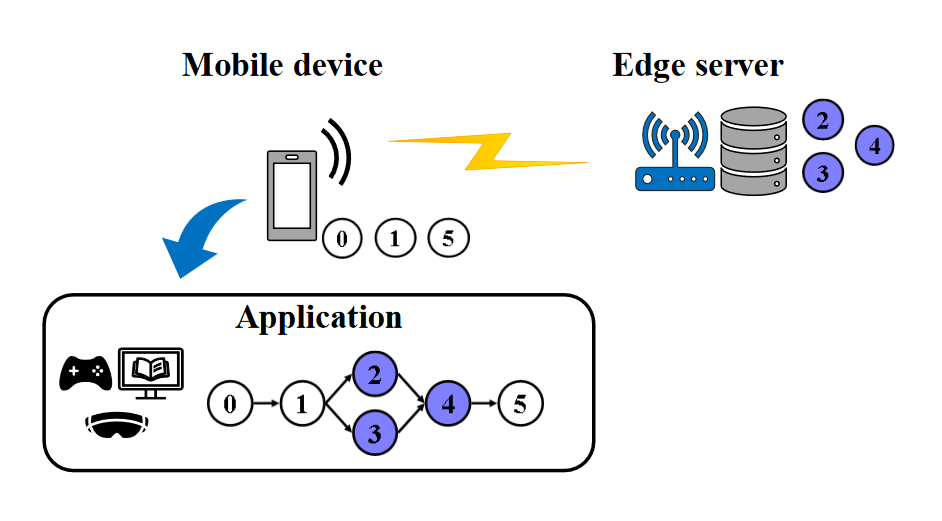}
  \caption{MEC system model in the DAPO problem.}
  \label{fig:system}
\end{figure}

The MEC system model in this paper is described by Fig.~\ref{fig:system}, which depicts a single-user scenario. Applications run on end-user mobile devices, such as smartphones. These applications can be of any nature~\citep{parvez2018}, such as video streaming, augmented reality, virtual reality, e-health, e-learning, etc. Furthermore, applications can be partitioned into tasks, sequential or parallel, which are individual pieces of code to be executed on the mobile device or sent for remote processing on a nearby server at the network edge. Here it is assumed that both the mobile device and the edge server are capable of fully executing the applications.

Data transmission uses the wireless radio channel between the mobile device and the edge server. Still referring to Fig.~\ref{fig:system}, the application in the example has six tasks, and a possible offloading decision is shown. Tasks~2,~3 and~4 are offloaded to the edge server, while tasks~0,~1 and~5 remain assigned to the mobile device for local processing. Data transmissions are performed whenever task processing is offloaded and can occur from mobile device to edge server or in the opposite direction. The data transmitted correspond to the input parameters required to run a task, such as variables, which are passed from a preceding task to a succeeding task.

\subsection{DAG application}
\label{subsec:3.1}

Applications can be described as DAGs~$G(V,E)$, where vertex set $V$ is the set of tasks and directed-edge set $E$ is the set of data transmissions (dependencies) between tasks. A task $i \in V$ can run on either the mobile device or the edge server. The number of parallel tasks that can be executed concomitantly, either on the mobile device or on the edge server, is not limited by the number of CPUs of each device. It is up to the scheduler of each machine to allocate CPUs, as the tasks can share the same CPU, which means that a CPU does not necessarily run the same task 100\% of the time. Also, with each task $i$ a binary offloading decision $o_i$ is associated, where $o_i = 0$ represents task execution on the mobile device and $o_i = 1$ represents execution on the edge server. The set $\{o_i, \forall i \in V\}$ is the offloading decision $D$ for this application. Since a task is considered an indivisible part of an application, its processing cannot be interrupted and then migrated from one device to the other: a task must run entirely on the mobile device or on the edge server. In the same way, it is not possible to run a task on two or more CPUs at the same time to speed up processing.

An edge $(i,j) \in E$ indicates that the data passed from task $i$ to task $j$ are input parameters necessary for the execution of task $j$. A task can only start being executed if all data from all its incoming edges are received. In the case of multiple outgoing edges, as soon as a task $i$ finishes its processing, it starts transmitting in parallel the corresponding data for each of those edges, requiring bandwidth sharing of the communication channel.

For the sake of simplicity, two assumptions are made. First, data transmissions cannot be interrupted and always finish successfully. Second, it is assumed that the underlying undirected graph is connected, and that there is only one task with no incoming edges (known as the initial task) and only one without outgoing edges (known as the ending task). These two tasks represent application start-up and result consumption. Both run necessarily on the end user's mobile device, as the application is started by the mobile device and the result of executing the application must be returned to it~\citep{shu2017}. Thus, neither the initial nor the ending task can be offloaded. Since the offloading decisions for these two tasks are already made in advance, the outgoing edges from the initial task and the incoming edges to the ending task are called anchors. The set of anchors is a subset of $E$.

Considering an application with $N$ offloadable tasks, tasks $t_0$ and $t_{N+1}$ are the initial and ending tasks, respectively. Application makespan is the total time elapsed from the beginning of $t_0$ to the end of $t_{N+1}$, when all tasks are done processing and there is no ongoing data transmission. Task processing times and data transmission times are defined next.

\subsection{Task processing time}
\label{subsec:3.2}

Task processing time is given by the time it takes a task to be fully executed on a single CPU. However, it may take longer if its allocated CPU is shared with other tasks. In fact, both the mobile device and the edge server can host more tasks running in parallel than their number of CPUs. A new task starts processing immediately after receiving all data from all its incoming edges, regardless of the number of tasks already running. It is up to the task scheduler to run them evenly.

When a task is allocated to a single CPU without sharing it with other tasks, the processing time of task $i$ is given by
\begin{equation}
t_i = \frac{n}{f c},
\label{eq:1}
\end{equation}
where $n$ is the task's number of operations, $f$ is the clock speed (in cycles/second), and $c$ is the CPU's computing capacity (in operations/cycle). However, by assuming CPU sharing, the processing time of a task increases as other tasks start running on the same CPU. If there are $u$ CPUs on a device, up to $u$ tasks can run in parallel without any sharing (100\% CPU occupancy for each task). As soon as a new task starts, all running tasks receive only a fraction of a CPU, extending their remaining processing time by the factor $(u + 1) / u$, as there are more tasks than CPUs available. Likewise, when a task finishes processing, each remaining task receives back a larger fraction of CPU occupancy, up to the maximum value of 100\%, which implies no more CPU sharing. Here, this instantaneous fraction is called a CPU scale factor $k_p$, defined by
\begin{equation}
  k_p = \text{max} \Big\{1,\frac{p}{u}\Big\},       
\label{eq:2a}
\end{equation}
where $p$ is the number of tasks currently running in parallel. Since a task can never be sped up by using more than one CPU, $k_p$ can never be less than 1, which means that the processing time given by Eq.~(\ref{eq:1}) is also the task's minimal completion time.

Taking Eq.~(\ref{eq:2a}) into account, the remaining processing time for $t_i$, denoted by $\Delta t_i$, is updated whenever the CPU scale factor varies (that is, a new task starts/finishes running on that device) and the task receives a smaller/larger CPU share. The new value for the remaining processing time $\Delta t'_i$ is given by
\begin{equation}
\Delta t'_i = \frac{\Delta t_i k'_p}{k_p},
\label{eq:2b}
\end{equation}
where $k_p$ represents the previous CPU scale factor, and $k'_p$ is the new one. Having $k'_p > k_p$ means that a new task has begun running on that device; $k'_p < k_p$ means a task has ended.

It is worth noting that, as the number of tasks running on each device varies, the processing time for a task on the mobile device can become lower than that on the edge server due to their CPU scale factors, even if the edge server has a more powerful hardware. Thus, offloading tasks to the edge server whenever possible may result in a longer makespan.

\subsection{Data transmission time}
\label{subsec:3.3}

The wireless radio channel transmission rate (in bytes/second) is denoted by $r$ and assumed to be the same for both uplink and downlink transmissions. Each edge $(i,j)$ carries $d_{i,j}$ bytes from task $i$ to task $j$. Thus, if there is no other ongoing transmission, the corresponding transmission time $T_{i,j}$ is given by
\begin{equation}
T_{i,j} = \frac{d_{i,j}}{r}.
\label{eq:3}
\end{equation}

However, by assuming a shared communication channel, multiple ongoing transmissions can take place. Channel bandwidth is shared evenly among the number of current data transmissions $k_t$, meaning all transmissions have the same priority. Whenever a new transmission starts or a previous transmission ends, the remaining time for all $d_{i,j}$ bytes to be transmitted, denoted by $\Delta T_{i,j}$, is updated. The new value for the remaining time $\Delta T'_{i,j}$ is given by
\begin{equation}
\Delta T'_{i,j} = \frac{\Delta T_{i,j} k'_t}{k_t},
\label{eq:4}
\end{equation}
where $ k'_t=k_t-1 $ if a transmission has ended or $ k'_t=k_t+1 $ if a new one has started.

\section{Heuristic approach}
\label{sec:4}

Providing an optimal solution to the DAPO problem in MEC is known to be difficult~\citep{shu2017}. Decisions based on integer linear optimization are present in the literature and they are proved to be NP-hard~\citep{mao2017,feng2022}. All known exact approaches are therefore excessively time-consuming~\citep{lin2020}, so a heuristic is needed.

\subsection{Overview}
\label{subsec:4.1}

The following list summarizes the key observations underlying the design of FlexDO.

\begin{itemize}
    \item Makespan reduction does not occur monotonically as more tasks are offloaded. Some offloading decisions, each one specifying where to run each task of a DAG application, can provide worse results than having no offloading at all. Even if offloading a task can shorten its processing time, this shortening can be offset by the time to transmit data from/to the task in question. For this reason, the proposed algorithm generates a set of offloading decisions, and tests each of them to select the one that results in the shortest makespan. As such, it does not rely on the last decision generated or on the decision that offloads the largest number of tasks.
    \item Some tasks are more relevant to makespan reduction than others, that is, the time difference (gain) between running a task locally or on the edge server varies from task to task in a DAG.
    \item Transmission times tend to weigh more heavily on makespan than processing times, but they can be canceled out. Task processing time only increases when there are more tasks than CPUs available (implying CPU sharing). However, as few as two transmissions in parallel are already enough for each one to have access to only half the available bandwidth (evenly shared), hence longer transmissions should probably be avoided by offloading the task pairs involved.
    \item The one-climb policy is not an absolute truth for the optimal offloading decision on a general DAG.
\end{itemize}

The one-climb policy is a recurring assumption in the literature~\citep{zhang2012,zhang2013,jia2014,mao2017,guo2019,yan2020}. This policy is based on the notion of a discontinuous offloading decision for a DAG application. If $D$ is such a decision, then there exists at least one path in the DAG along which at least two tasks are marked for offloading by $D$ while at least one other lying between them is not. That is, task processing goes down to the mobile device and up again to the edge server within that path. The one-climb policy states that, if $D$ is a discontinuous offloading decision, then there will always be at least one continuous offloading decision $D'$ resulting in a lower application makespan than $D$. Decision $D'$, therefore, has at most one uplink transmission and one downlink transmission on every path in the DAG. Its proof for sequential DAG applications can be found in~\citet{yang2016}. 

Although the one-climb policy does hold for linear DAGs, in the general case it does not because of CPU limitations on the edge server and parallelism. It is possible for an edge server to be overloaded to the point where it is faster to process the task on the mobile device. Furthermore, when an offloading decision is forced to become continuous, it may require new data transmissions involving the tasks that were thus offloaded. These additional data transmissions can have an impact on the makespan.

\begin{figure}[t]
  \centering
  \includegraphics[width=0.8\textwidth]{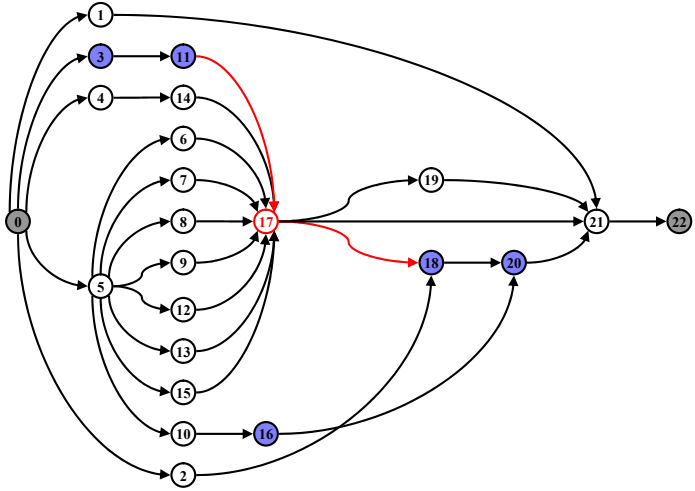}
  \caption{Implications of the one-climb policy for general DAGs.}
  \label{fig:oneclimb2}
\end{figure}

One of the DAG applications from the data set used in this paper is shown in Fig.~\ref{fig:oneclimb2}. Tasks in blue are the ones that should be offloaded to the edge server in the optimal offloading decision, which results in a makespan of 139.17~s. However, there is a discontinuity in a single DAG path, highlighted in red in the figure. When offloading task~17, the offloading decision becomes continuous, complying with the one-climb policy and canceling the transmission time of edges~(11, 17) and~(17, 18). However, this reduction negatively affects the total makespan of the DAG application, as it now performs the data transmissions on edges~(6, 17), (7, 17), (8, 17), (9, 17), (12, 17), (13, 17), (14, 17), (15, 17), (17, 19), and (17, 21). The makespan of this continuous solution is 237.04~s, a deterioration of ~70\% in relation to the optimal decision. Thus, a discontinuous offloading scheme is not necessarily a bad decision. In fact, the results in Section~\ref{sec:6} show that around 17\% of the optimal decisions for our data violate the one-climb policy. 

\subsection{FlexDO}
\label{subsec:4.2}

FlexDO generates a small set of offloading decisions to reduce the makespan of an application, beginning with the decision to offload no tasks and in the end testing each decision and choosing the one that provides the shortest makespan. This is done in two phases. In the first phase, it greedily searches for the edge with the longest data transmission time and offloads its two end tasks to obviate the data transmission between them. An important detail here is that if an edge shares an end task with an anchor, the transmission time with which it participates in this first phase is its original transmission time minus that of the anchors. This process continues on to the next edge with longest transmission time, now prioritizing edges with an end task already marked for offloading. If no edges remain connecting an offloaded task to a non-offloaded task, the edge with the longest transmission time is selected. As each new edge is selected in this sequence, it joins the previous ones in forming another offloading decision for later consideration.

For a DAG application with $N$ tasks, the first phase continues until $s_I$ tasks have been marked for offloading, where $s_I < N$ is an input parameter. The second phase of FlexDO will then generate all $2^{s_{II}}$ possibilities of offloading decisions for the remaining $s_{II} = N - s_I$ tasks, while keeping the offloading marks from the first phase. After the second phase, FlexDO calculates the makespan associated with each decision and selects the one that results in the shortest makespan.

\begin{figure}[p]
  \centering
  \subfloat[a][A general DAG application.]{\includegraphics[width=0.7\textwidth]{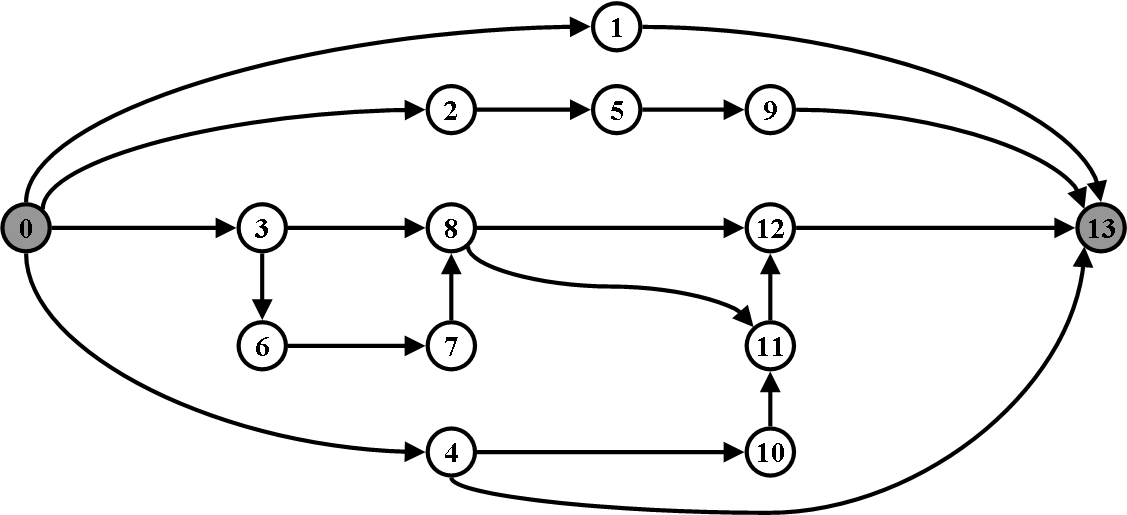} \label{fig:heuristic-a}} \\
  \subfloat[b][Offloading decisions generated in each phase.]{\includegraphics[width=0.8\textwidth]{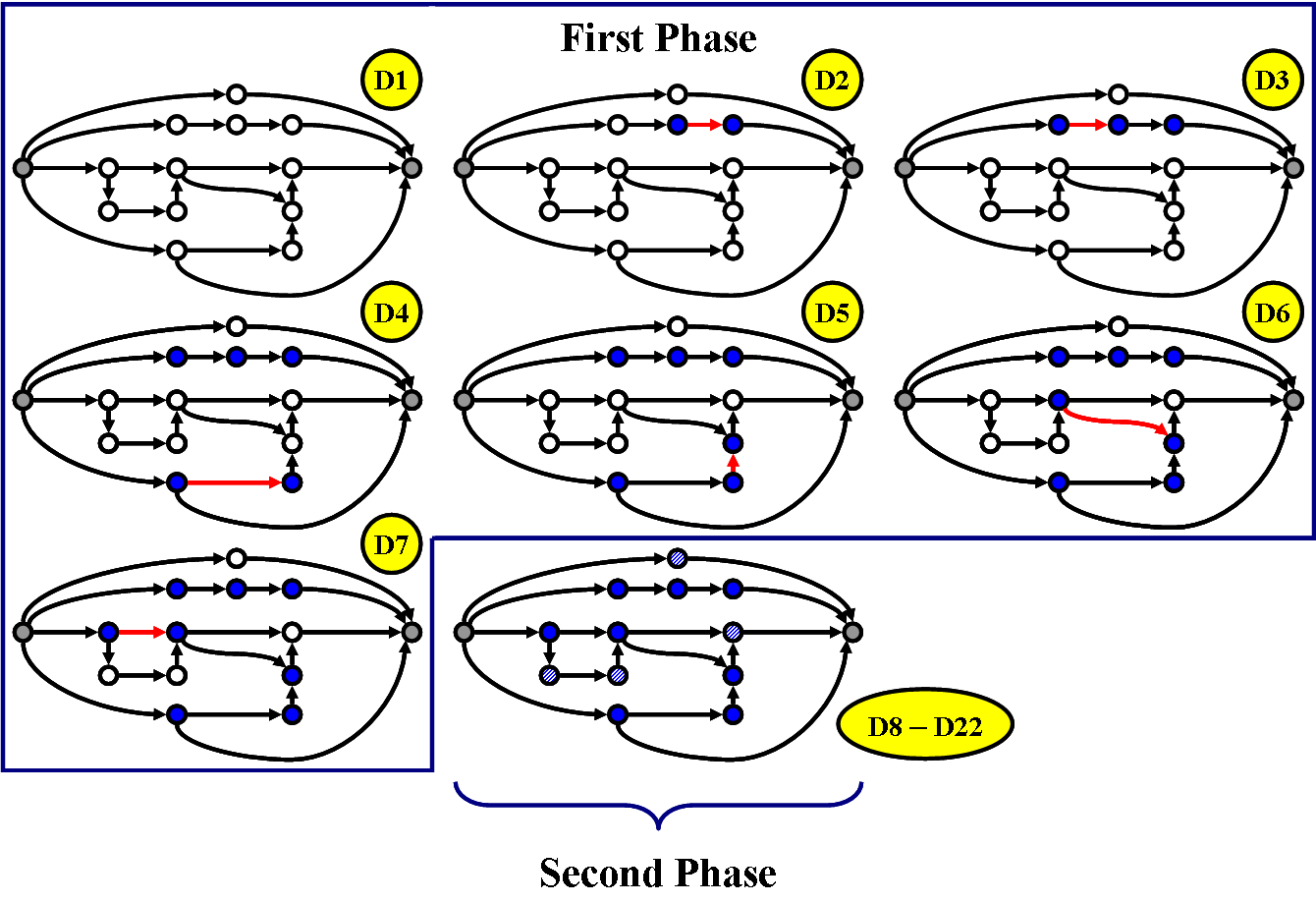} \label{fig:heuristic-b}}
  \caption{An example of how FlexDO operates.}
  \label{fig:heuristic}
\end{figure}

An example of how the FlexDO heuristic works is shown in Fig.~\ref{fig:heuristic}. White tasks are marked to run on the mobile device and blue tasks to be offloaded to the edge server. Since the first and the last tasks always run on the mobile device and cannot be offloaded, they are grayed out. This example is for a DAG application with~12 offloadable tasks and $s_I=8$; that is, the first phase lasts until~8 tasks have been marked to be offloaded.

The first offloading decision generated by FlexDO~(D1) is the No Offloading decision. Now suppose edge~(5, 9) has a transmission time of 80~s, the largest in the application. Since task~9 is anchored to task~13, the transmission time of edge~(9, 13) must be subtracted from that of edge~(5, 9). If data transmission on edge~(9, 13) takes 12~s, then the time associated with edge~(5, 9) for FlexDO operation is 68~s. Assuming this is still the most time-consuming transmission over all edges, tasks~5~and~9 are both marked for offloading~(D2). Now the only available edge connected to tasks~5~or~9 is edge~(2, 5), resulting in task~2 being marked for offloading~(D3). From this point on, assume the first phase generates four more decisions~(D4--D7), marking tasks~4, 10, 11, 8, and~3 for offloading.

At this point, FlexDO takes on its second phase and generates additional $2^4 = 16$ other offloading decisions in regard to tasks~1, 6, 7, and~12. First-phase~D7 is one of them, so only~15 are actually generated. The makespan of each of D1--D22 is calculated, and the one yielding the shortest makespan is output by FlexDO.

Note that it is possible that there are not enough edges to go through in the first phase until the threshold $s_I$ is reached. This happens when the DAG has many tasks associated only with anchor edges, as in the case of task~1 in the example. When this happens, an offloading gain is calculated for each task not yet marked for offloading, given by
\begin{equation}
G_{k} = t^m_{k} - t^e_{k} - \sum_{\substack{(i,j) \in E \mid \\ k \in \{i,j\}}} T_{i,j}
\label{eq:4a}
\end{equation}
for task $k$, where $t^m_{k}$ and $t^e_{k}$ are the processing times of task $k$ on the mobile device and on the edge server, respectively, and $T_{i,j}$ is the transmission time of edge $(i,j)$, provided $k=i$ or $k=j$. Then, tasks are marked for offloading, from the highest gain to the lowest, which generates additional offloading decisions, until the threshold $s_I$ is reached and the second phase begins. Pseudocode for FlexDO is given in Algorithm~\ref{alg:heuristic}.

\begin{algorithm}[ht]
\footnotesize
\SetAlgoLined
\DontPrintSemicolon
\SetAlgorithmName{Algorithm}{Algorithm}{}
\SetKwInput{KwInput}{input}
\SetKwInput{KwOutput}{output}
\textbf{let} $L$ be the set of non-anchor edges in $E$\;
\For{$(i,j)\in L$}{
    \textbf{decrease} $T_{i,j}$ by every $T_{u,v}$ such that
$(u,v)\in E\setminus L$ with $u\in\{i,j\}$ or $v\in\{i,j\}$\;
}
\textbf{let} $D$ be the offloading decision in which no task is marked for offloading\;
$\mathcal{D}\leftarrow\{D\}$\;
\While{$L\neq\emptyset \textup{ and fewer than } s_I \textup{ tasks are marked for offloading in } D$}{
    \If{$\textup{there exists } (i,j) \in L \textup{ such that } \ell \textup{ is the only task in } \{i,j\} \textup{ not marked}$ $\textup{for offloading in } D$}{
        \textbf{mark} $\ell$ for offloading in $D$, provided $T_{i,j}$ is longest over all such edges\;
    }
    \Else{\textbf{mark} both $i$ and $j$ for offloading in $D$, provided $T_{i,j}$ is longest over $L$}
    $\mathcal{D}\leftarrow\mathcal{D}\cup\{D\}$\;
    $L\leftarrow L\setminus\{(i,j)\}$\;
}
\textbf{let} $P$ be the set of tasks not marked for offloading in $D$\;
\While{$\textup{fewer than } s_I \textup{ tasks are marked for offloading in } D$}{
    \textbf{mark} task $k$ for offloading in $D$, provided $G_k$ is greatest over $P$\;
    $\mathcal{D}\leftarrow\mathcal{D}\cup\{D\}$\;
    $P\leftarrow P\setminus\{k\}$\;
}
\For{$\textup{each of the } 2^{s_{II}} \textup{ possible joint markings for offloading of the } s_{II} \textup{ tasks not}$ $\textup{marked for offloading in } D$}{
    $D'\leftarrow D$\;
    \textbf{mark} the $s_{II}$ tasks for offloading in $D'$ according to the joint marking being considered\;
    $\mathcal{D}\leftarrow\mathcal{D}\cup\{D'\}$\;
}
\Return{\textup{the decision of shortest makespan in} $\mathcal{D}$}\;
\caption{FlexDO.}
\label{alg:heuristic}
\end{algorithm}

\subsection{Computational complexity}
\label{subsec:4.3}

The time complexity of Algorithm~\ref{alg:heuristic} can be broken up into four components.

\begin{enumerate}
    \item The sorting of edges (by transmission times) and vertices (by gains), which is implicitly assumed. Considering that $\vert L\vert$ is $O(N^2)$, all the necessary sorting can be done in $O(N^2\log N)$ time.
    \item The loops in lines~6 and~14, taken together, iterate $s_I=O(N)$ times, each iteration requiring $O(1)$ time to complete.
    \item The loop in line~18, which iterates $2^{N-s_I}$ times, spends $O(N)$ in each iteration. For $N-s_I=O(\log N)$, as in Section~\ref{sec:6}, this loop spends $O(N^2)$ time.
    \item The makespan calculations in line~22, one for each of $N$ offloading decisions, can be carried out for each decision as follows. Traverse the DAG breadth-first, in $O(N^2)$ time, and build a length-$O(N^2)$ event queue along the way. The events in this queue are the intertwined start and stop events of task executions and data transmissions. Sorting this queue by those events' time tags in ascending order in $O(N^2\log N)$ time, and then processing it in $O(N^2)$ time, yields the DAG's makespan. Overall, $O(N^3\log N)$ time is needed.
\end{enumerate}

Adding up all four contributions yields an $O(N^3\log N)$ time complexity, which is slightly higher than the $O(N^3)$ reported for HOA~\citep{guo2019}. Of course, for DAGs with $\vert L\vert=O(N\log N)$ this gets reduced to $O(N^2\log^2 N)$, and for $\vert L\vert=O(N)$ it goes further down to $O(N^2\log N)$.

\section{Data analysis}
\label{sec:5}

FlexDO was developed to perform independently of DAG topology, based on realistic data. Most recent works deal with sequential~\citep{ning2019,wang2021}, tree-based~\citep{duan2021}, or fan-in/fan-out DAGs~\citep{mahmoodi2019,mazouzi2019}. Others do not even consider dependencies between the tasks of an application, targeting instead other aspects of the DAPO problem~\citep{tran2019,peng2021}, such as atomic tasks modeled as sequences of bits submitted in batch.

For the present study, the DAG applications were extracted from the Alibaba Cluster Trace Program from 2018.\footnote{~https://github.com/alibaba/clusterdata} This data set contains information from batch job tasks for millions of DAG applications submitted to a cluster of more than 4\,000 machines.

\subsection{Data set description}
\label{subsec:5.1}

The data set describes its DAG applications in two files, one for the applications, which includes DAG topology, and another for instances, which have the internal specification of the tasks. These files contain information about memory size, percentage CPU utilization, start time, and end time of each instance, for example. As the instances are part of a task (i.e., a DAG vertex), and further details about their inner dependencies are omitted from the data set files, we summarize all instances for a given task as a single instance of processing time given by the average time of all instances.

Exact CPU and memory performance figures are also missing from the data set, but hardware information can be assumed based on Alibaba Cloud's instance family SCCG5, a general-purpose supercomputing cluster.\footnote{~https://www.alibabacloud.com/help/en/doc-detail/25378.htm\#sccg5} Table~\ref{tab:cluster} shows the values used in this paper. Additional information about floating-point operations~(FLOPs) per cycle is available in~\citet{dolbeau2018}.

\begin{table}[t]
\caption{Hardware information for SCCG5 machines.}
\label{tab:cluster}
\resizebox{\textwidth}{!}{
\begin{tabular}{||c|c|c|c|c||}
\hline
CPU architecture & Clock & No. of CPUs & RAM & FLOPs/cycle\footnotesize{*} \\ \hline \hline
Intel Xeon Platinum 8163 & 2.5~GHz & 48 & 384~GiB & 32 \\ \hline
\end{tabular}
}
\begin{tablenotes}[flushleft]
   \item \footnotesize{* https://www.cpu-world.com/CPUs/Xeon/Intel-Xeon\%208163.html}
\end{tablenotes}

\end{table}

Although there is information about how much memory is used by each task, it is not possible to infer from the data set how many bits are passed in data transmissions between tasks. Thus, here it is assumed that the total amount of data passed on all edges incoming to a given task is chosen randomly up to the total size of the data structure used by that task.

\subsection{DAG preparation}

\label{subsec:5.2}

The Alibaba data set contains more than 4 million DAG applications. We selected 1\,322 of them from different parts of the data set, searching it while abiding by the following criteria.

\begin{itemize}
    \item Include DAGs that have 18 to 28 tasks. The lower bound removes DAGs with overly simple topologies and the upper bound is applied so that it is still reasonable to find the optimal offloading decision through exhaustive enumeration.
    \item Remove DAGs with incomplete topology information.
    \item Remove DAGs without memory or CPU usage information.
    \item Keep only DAGs that have run successfully to completion.
    \item Interrupt the search after it has run for an hour.
\end{itemize}

\begin{figure}[t]
  \centering
  \includegraphics[width=0.6\textwidth]{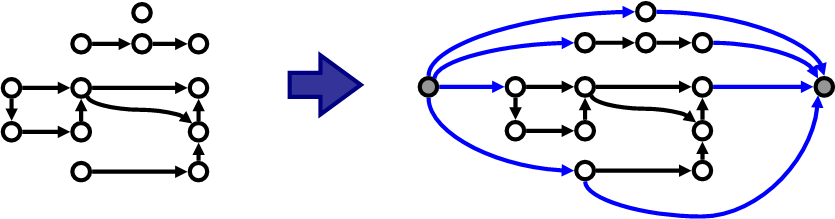}
  \caption{Addition of the two extra tasks and their associated edges (in blue).}
  \label{fig:dummy}
\end{figure}

Knowing that the data set includes DAG applications running on Alibaba clusters, it is reasonable to assume that these DAGs were submitted by users and that the results were returned to them. Thus, after obtaining the 1\,322 DAGs, two tasks are added to each DAG to mimic the behavior of an application running on a mobile device. An initial task is connected to the tasks that do not have incoming edges, and an ending task is connected to those that do not have outgoing edges, as shown in Fig.~\ref{fig:dummy}. Neither task can be offloaded, hence they are both assigned zero processing time (any nonnegative constant would do here, as it affects the makespan of every DAG by the same amount). Their transmission times are generated as described in Section~\ref{subsec:5.1}.

\subsection{DAG properties}

\begin{figure}[p]
  \centering
  \subfloat{\includegraphics[width=0.65\textwidth]{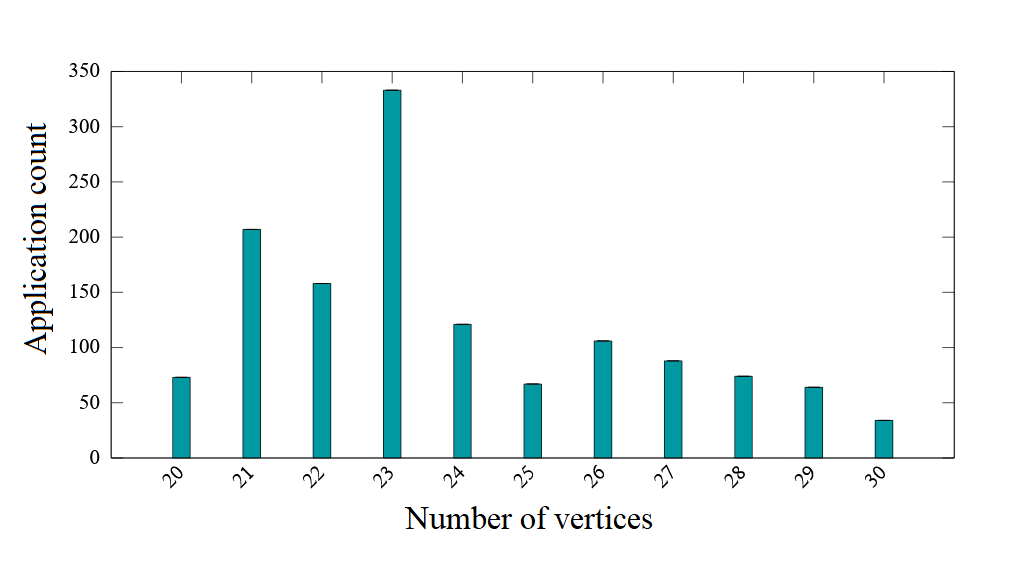} \label{fig:dags-a}} \\
  \subfloat{\includegraphics[width=0.65\textwidth]{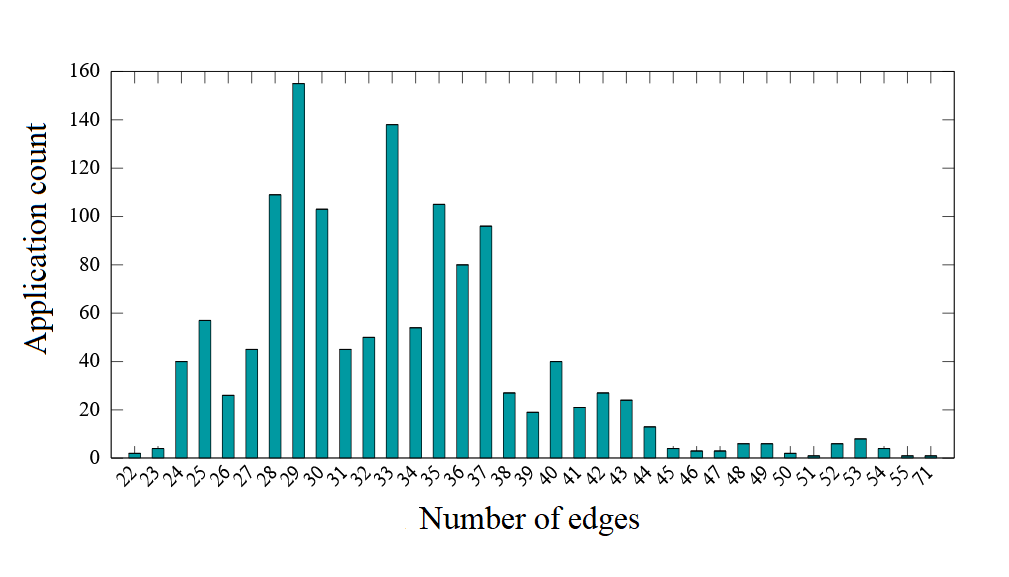} \label{fig:dags-b}} \\
  \subfloat{\includegraphics[width=0.65\textwidth]{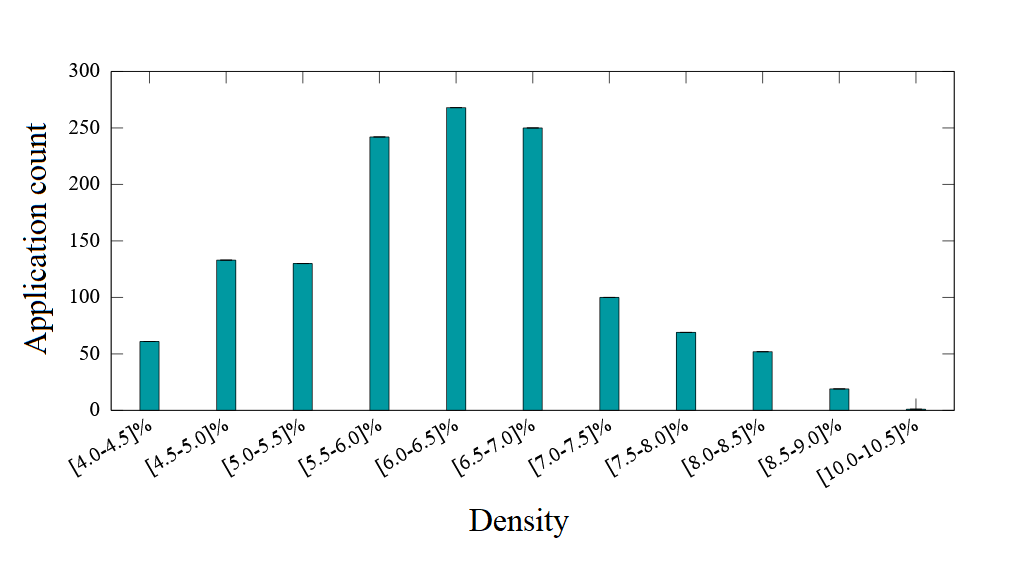} \label{fig:dags-c}}
  \caption{Distributions of DAG properties.}
  \label{fig:dags}
\end{figure}

In addition to having realistic DAG applications, it is also important that DAGs be diverse in their structure, as the FlexDO heuristic is expected to perform well in any scenario. Figure~\ref{fig:dags} shows how the DAG applications are distributed according to number of vertices, number of edges, and density. Density varies between 0, when a graph has no edges, and 1, when there is an edge for each of the $\frac{N(N-1)}{2}$ unordered pair of vertices.

As can be seen in the figure, the most common value of $N$ is 23, which amounts to a little less than 25\% of all DAGs. The remaining 75\% are spread out less unevenly. As for the DAGs' numbers of edges and densities, these seem to thin out at the lower and higher values. This is reasonable, as these extremes generally bespeak little structured DAG: too few dependencies among tasks in the former case, too many in the latter.

\section{Results}
\label{sec:6}

To evaluate the performance of FlexDO, three scenarios are used. The first scenario considers the variation in the number of CPUs available on the mobile device. The second considers this variation on the edge server. The third considers different transmission rates for the shared communication channel. In each of these scenarios, FlexDO is tested for three different values of $s_{II}$, namely, $s_{II} \in \{0,\log_2N,\log_2N^2\}$, following $s_{II} = N - s_I$. These values for $s_{II}$ generate $1$, $N$, or $N^2$ additional offloading decisions, respectively, in the second phase of Algorithm~\ref{alg:heuristic}. Throughout these experiments, FlexDO is compared with the Heuristic Offloading Algorithm (HOA) from~\citet{guo2019}, and with two baseline strategies (Full Offloading and No Offloading).


HOA always assumes an unlimited number of CPUs on the mobile device. If all CPUs on the edge server are busy, new tasks are queued until a CPU becomes available, rather than sharing CPUs with the tasks already running. Also, HOA does not perform more than one data transmission at a time, scheduling transmissions according to a FIFO scheme. These differences between the model underlying FlexDO and that underlying HOA imply that optimal decisions can be different for the same DAG. We follow~\citet{guo2019} faithfully as far as implementing and evaluating HOA are concerned.

For the sake of fairness, results are therefore presented in terms of the relative difference between the heuristic makespan $M_\mathrm{heur}$ and the optimal makespan $M_\mathrm{opt}$. FlexDO and the baseline strategies follow the modeling described in Section~\ref{sec:3} and are compared with the corresponding optimal makespans. The HOA results are relative to the optimal makespans obtained within the system modeling of~\citet{guo2019}.

Since the processing times of the Alibaba DAG applications refer to a large-scale computing cluster, these times are modified to better fit the hardware of commercial edge servers and current smartphones, whose characteristics are shown in the Table~\ref{tab:hardware}, according to
\begin{equation}
t^m_{k} = t^e_{k} \Big(\frac{f_e c_e}{f_m c_m}\Big)
\label{eq:6}
\end{equation}
and
\begin{equation}
t^e_{k} = t^c_{k} \Big(\frac{f_c c_c}{f_e c_e}\Big),
\label{eq:5}
\end{equation}
where $t^m_{k}$, $f_m$, and $c_m$ are the processing time of task $k$ in seconds, clock speed in Hz, and computing capacity in operations/cycle at the mobile device, respectively; $t^e_{k}$, $f_e$, and $c_e$ are for the edge server; and $t^c_{k}$, $f_c$, and $c_c$ are for the computing cluster.

\begin{table}[t]
\caption{Hardware information for the mobile device and the edge server.}
\label{tab:simulation}
\resizebox{\textwidth}{!}{
\begin{tabular}{||c|c|c|c|c||}
\hline
Type & CPU architecture & Clock & No. of CPUs & FLOPs/cycle\footnotesize{*} \\ \hline \hline
Mobile device\footnotesize{**} & Qualcomm Snapdragon 865 & 1x2.84, 3x2.42, 4x1.8~GHz & 8 & 8 \\ \hline
Edge server\footnotesize{***} & Intel Xeon D-2100 & 2.0~GHz & 16 & 32 \\ \hline
\end{tabular}
}
\begin{tablenotes}[flushleft]
   \item \footnotesize{* https://en.wikichip.org/wiki/flops\#x86}
   \item \footnotesize{** https://www.qualcomm.com/products/mobile/snapdragon/smartphones/snapdragon-8-series-mobile-platforms/snapdragon-865-5g-mobile-platform}
   \item \footnotesize{***~https://www.intel.com/content/www/us/en/products/docs/processors/xeon/d-2100}
   \item \footnotesize{-brief.html}
\end{tablenotes}
\label{tab:hardware}
\end{table}

\subsection{Varying the number of CPUs of the mobile device}
\label{subsec:6.1}

A mobile device may not have all its CPUs available for processing a DAG application, dedicating only part of them to it. This scenario assumes the availability of 2, 4, 8, or an unlimited number of CPUs on the mobile device, and the availability of all 16 CPUs of the edge server. The transmission rate of the wireless channel is assumed to be 20 Mbps, which is common in legacy Long Term Evolution~(LTE) networks~\citep{saliba2019}.

\begin{figure}[t]
  \centering
  \includegraphics[width=0.8\textwidth]{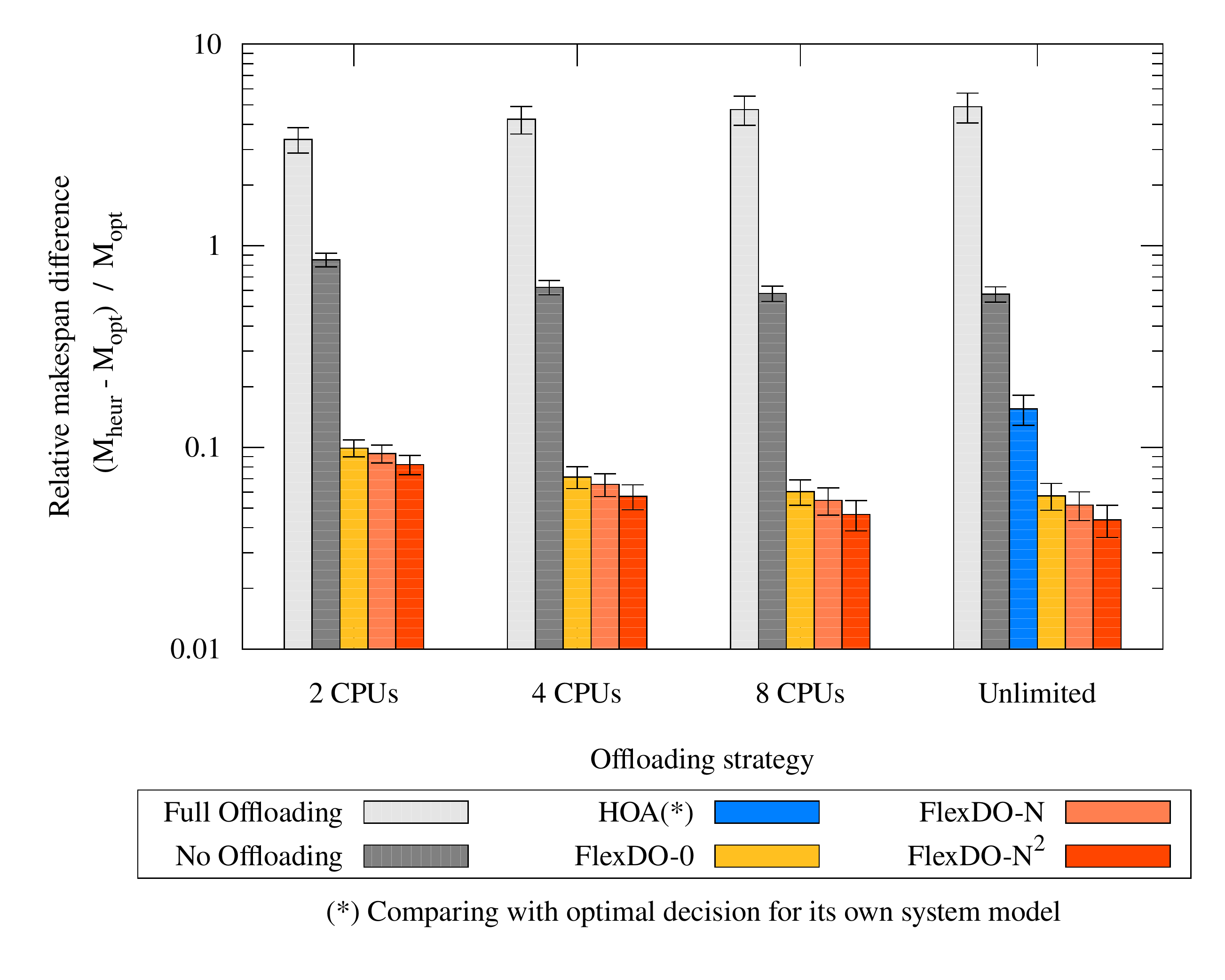}
  \caption{Difference in relative makespan due to variation in the number of CPUs on the mobile device.}
  \label{fig:localcpu}
\end{figure}

Figure~\ref{fig:localcpu} shows that FlexDO's performance remains stable regardless of the number of CPUs on the mobile device. FlexDO can only be meaningfully compared to HOA when both consider an unlimited number of CPUs at the mobile device, with FlexDO presenting results ranging from 4.3\% to 5.7\% above the optimal makespan, while HOA is 15.5\% above its own optimal solution. It is worth noting that the fact that HOA always considers unlimited resources makes its makespan calculations optimistic, serving as an upper bound for the real makespan. And yet, even FlexDO-0, which tests a very limited set of offloading decisions, performs better. As shown in Fig.~\ref{fig:localcpu}, the baseline solutions lead to unacceptably high makespans, since these strategies do not take into account task processing times or transmission times.

Still referring to Fig.~\ref{fig:localcpu}, it must be noted that only in the Full Offloading case does the performance worsen as the number of CPUs increases on the mobile device. This is due to the fact that the increased number of CPUs does not benefit the Full Offloading decision, but rather only reduces the makespan of the optimal solution.

\subsection{Varying the number of CPUs of the edge server (multi-user case)}
\label{subsec:6.2}

This scenario investigates the impact on makespan of the number of CPUs available on the edge server. In a scenario with multiple mobile devices associated with a single edge server, a reduced number of CPUs at the edge may be available to process the DAG applications. This reduction is therefore a useful proxy to emulate the multi-user case.

Here, an unlimited number of CPUs on the mobile device is assumed for compatibility with HOA. On the edge server side, there are 1, 2, 4, or 16 CPUs (out of the 16 existing CPUs) available to a user. The transmission rate is kept 20~Mbps, as before.

\begin{figure}[t]
  \centering
  \includegraphics[width=0.8\textwidth]{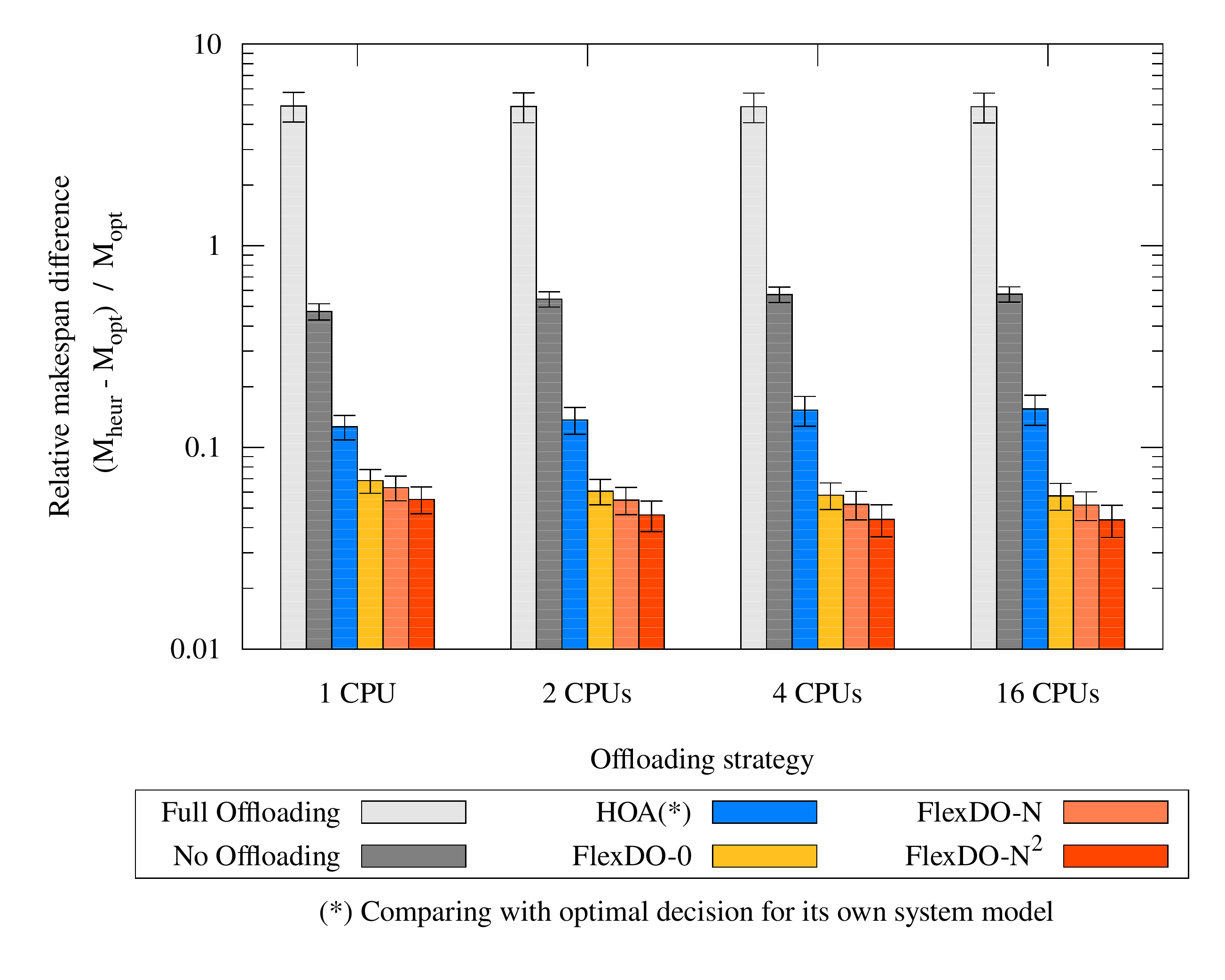}
  \caption{Difference in relative makespan due to variation in the number of CPUs on the edge server.}
  \label{fig:edgecpu}
\end{figure}

Figure~\ref{fig:edgecpu} shows the consistency of the three FlexDO versions, which are 5.5\% to 6.8\% above the optimal makespan when only one CPU is available on the edge server. These results become even better as more CPUs become available, achieving a makespan value only 4.3\% to 5.7\% above the optimal when all 16 CPUs are available. On the other hand, HOA's performance worsens relative to its optimal solution as more CPUs are available, going from 12.6\% above the optimal makespan in the single-CPU case to 15.5\% in the 16-CPU case. This contrast is due to the fact that FlexDO is more successful at choosing an offloading decision by testing a set of decisions, and insisting on the offloading process instead of going back to a baseline solution such as No Offloading. Also, both baseline solutions perform significantly worse than HOA or FlexDO.

In addition to the relative makespan results, it is worth delving deeper into the absolute makespan for this multi-user case. Table~\ref{tab:absolute} shows the average makespan for all 1\,322 DAG applications. Since these DAGs are quite diverse, ranging from quite simple to processing/transmission-intensive applications, a high deviation is expected. However, it is possible to see that, on average, FlexDO achieves better makespan values than HOA even with a smaller number of CPUs on the edge server. With 16 CPUs, HOA has an average makespan of 348.53~s, while FlexDO reaches 333.78 to 341.95~s with as few as two CPUs.

\begin{table}[t]
\centering
\caption{Absolute makespan for HOA and FlexDO in the multi-user scenario.}
\resizebox{\textwidth}{!}{
\begin{tabular}{c|c|c|c|c||}
\cline{2-5}
& \multicolumn{1}{c|}{1 CPU} & \multicolumn{1}{c|}{2 CPUs} & \multicolumn{1}{c|}{4 CPUs} & 16 CPUs \\ \hline \hline
\multicolumn{1}{||c|}{HOA} & 385.95 $\pm$ 41.48 & 354.52 $\pm$ 36.49 & 349.05 $\pm$ 35.73 & \textbf{348.53 $\pm$ 35.68} \\ \cline{1-1} \hline
\multicolumn{1}{||c|}{FlexDO-0} & 370.60 $\pm$ 40.33 & \textbf{341.95 $\pm$ 36.72} & 334.12 $\pm$ 35.86 & 333.31 $\pm$ 35.78 \\ \cline{1-1} \hline
\multicolumn{1}{||c|}{FlexDO-N} & 366.27 $\pm$ 39.35 & \textbf{337.67 $\pm$ 35.91} & 329.77 $\pm$ 35.04 & 329.05 $\pm$ 34.96 \\ \cline{1-1} \hline
\multicolumn{1}{||c|}{FlexDO-N$^2$} & 361.53 $\pm$ 38.55 & \textbf{333.78 $\pm$ 35.48} & 326.37 $\pm$ 34.70 & 325.79 $\pm$ 34.65 \\ \hline
\end{tabular}
}
\label{tab:absolute}
\end{table}

\subsection{Varying the data transmission rate}
\label{subsec:6.3}

Transmission rates may be reduced due to channel conditions~\citep{mao2017}, legacy equipment, and high demand in a crowded cell. On the other hand, higher transmission rates are expected in 5th Generation (5G) networks. So variation in data transmission has a potential to impact the offloading decision process. Here, an unlimited number of CPUs on the mobile device is considered while all~16 edge server CPUs are available to the DAG application. The transmission rates are 10 Mbps (degraded/crowded channel), 20 Mbps (legacy), and 100 Mbps (end-user data rate for 5G as in~\citeauthor{parvez2018},~\citeyear{parvez2018}).

\begin{figure}[t]
  \centering
  \includegraphics[width=0.8\textwidth]{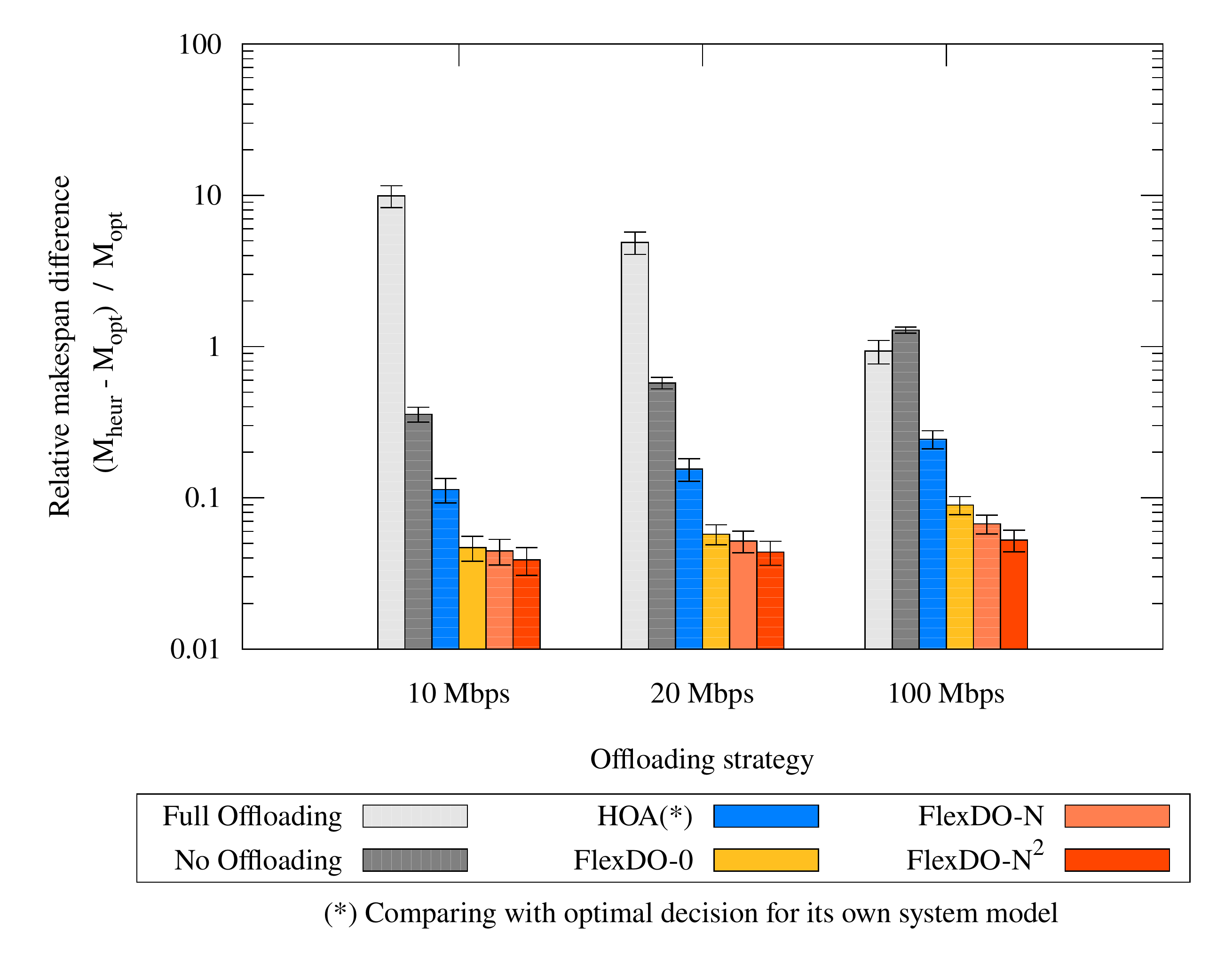}
  \caption{Difference in relative makespan due to variation in data transmission rate.}
  \label{fig:tx}
\end{figure}

The transmission-avoidance aspect of FlexDO provides the most benefit from different transmission rates. Since FlexDO prioritizes offloading task pairs in order to nullify higher data transmissions between them, edges with fewer data to be transmitted are good candidates for migrating task processing from/to the edge server. Figure~\ref{fig:tx} shows how close FlexDO gets to the optimal solution. For a rate of 10~Mbps, FlexDO is 3.9\% and 4.7\% above the optimal makespan. With 100~Mbps, this performance remains stable, being only 5.2\% to 8.9\% above the optimal makespan. For HOA, on the other hand, the tendency is to move further and further away from the optimal makespan as the transmission rate increases, deviating by up to 24.4\% from the optimal makespan with a transmission rate of 100~Mbps.

Considering the evolution of future networks in terms of transmission rate, a greater separation from the optimal solution as the rate increases may lead to poor offloading decisions by HOA. This behavior is due to the fact that HOA seeks a monotonically decreasing makespan reduction, and then resorts to the No Offloading decision immediately when this monotonic reduction does not occur.

\section{Concluding remarks}
\label{sec:7}

Given the results, it is clear that limiting FlexDO to neither the one-climb policy nor the monotonic reduction of makespan, and moreover generally not offloading tasks one by one, are beneficial in terms of obtaining average makespans lower than those of HOA in all scenarios. Furthermore, even though FlexDO does not calculate the impact of parallel transmissions or concurrent processing while generating decisions, these factors are duly considered when testing a small set of offloading decisions and selecting the decision that provides the shortest makespan. Also, FlexDO tries to offload those tasks that contribute most to reducing makespan, which has to do with the difference in processing time on the mobile device and on the edge server. It also cancels the highest data transmission times by keeping task pairs together on either the mobile device or the edge server. All these features refer back to the design principles behind FlexDO in Section~\ref{sec:4}, highlighting the importance of, as far as possible, considering all aspects that can possibly influence performance.

\section*{Declaration of competing interests}

The authors declare that they have no known competing financial
interests or personal relationships that could have appeared to influence
the work reported in this paper.

\section*{Acknowledgments}

We acknowledge partial support from Conselho Nacional de Desenvolvimento Cient\'\i fico e Tecnol\'ogico (CNPq), Coordena\c c\~ao de Aperfei\c coamento de Pessoal de N\'\i vel Superior (CAPES), and a BBP grant from Funda\c c\~ao Carlos Chagas Filho de Amparo \`a Pesquisa do Estado do Rio de Janeiro (FAPERJ). This work was also supported by MCTIC/CGI.br/São Paulo Research Foundation (FAPESP) through projects Slicing Future Internet Infrastructures (SFI2) – grant number 2018/23097-3, Smart 5G Core And MUltiRAn Integration (SAMURAI) – grant number 2020/05127-2 and Programmable Future Internet for Secure Software Architectures (PROFISSA) - grant number 2021/08211-7.

\appendix

\bibliographystyle{elsarticle-harv}
\bibliography{cas-refs}





\end{document}